\documentclass[aps,showpacs,twocolumn,prd,superscriptaddress]{revtex4}
\usepackage{epsfig,amsmath}

\begin{document}

\title{Stability of equatorial circular geodesics in static axially symmetric spacetimes}

\author{Guillermo A. Gonz\'alez}
\email{guillego@uis.edu.co}
\affiliation{Escuela de F\'{\i}sica, Universidad Industrial de Santander, A. A.
678, Bucaramanga, Colombia}

\author{Framsol L\'opez-Suspes} 
\email{framsol@gmail.com}
\affiliation{Escuela de F\'{\i}sica, Universidad Industrial de Santander, A. A.
678, Bucaramanga, Colombia}
\affiliation{Facultad de Telecomunicaciones, Universidad Santo Tom\'as,
Bucaramanga, Colombia}


\begin{abstract}
A general study of the stability of equatorial circular orbits in static axially
symmetric gravitating systems is presented. Important circular geodesics as the
marginally stable orbit, the marginally bounded orbit and the photon orbit are
analyzed. We found general expressions for the radius, specific energy, specific
angular momentum and the radius of the marginally stable orbit, both for null
and timelike circular geodesics. Solutions expressed in cylindrical coordinates,
oblate spheroidal coordinates, and prolate spheroidal coordinates are
considered. We show that all null circular orbits are unstable and that there are
not marginally stable null geodesics, whereas that for timelike geodesics the
orbits can be unbounded, bounded or circulars.
\end{abstract}

\maketitle

\section{Introduction}

A major problem in the General Theory of Relativity is obtaining exact solutions
of the Einstein equations corresponding to the gravitational field of acceptable
configurations of matter. Now then, as one of the most fundamental
characteristic of isolated systems in universe is axial symmetry, static or
stationary axially symmetric exact solutions are of great astrophysical
relevance. Accordingly, through the years, a great deal of work has been
dedicated to the theoretical study of this kind of exact solutions \cite{KSHM}.

On the other hand, once the solutions are found, another equally important
problem, closely related with the above study, is the analysis of the motion of
test particles in the gravitational field generated by such distributions of
matter. Indeed, the study of the motion of test particles provides valuable
information about the structure and behavior of such gravitational fields.
Furthermore, the study of orbits in the equatorial plane is of clear
astrophysical relevance due to its relation with the dynamics of intergalactic
stellar motion or the flow of particles in accretion disks around black holes.

The motion of test particles in axially symmetric spacetimes has been studied by
different authors through the years, both for satic as for stationary spacetimes
and with different configurations of sources (see, for instance, references
\cite{BAR} to \cite{DPS}). Now, the purpose of the preset work is a general
study of the stability of circular orbits in the equatorial plane in different
gravitating systems formed by axially symmetric structures. In particular, we
will analyze some important circular geodesics as the marginally stable orbit,
the marginally bounded orbit and the photon orbit.

The paper is organized as follows. Section \ref{sec:mot} is devoted to derive
the geodesic equations, the effective potential, and general expressions for the
main characteristic of circular orbits: the radius, specific energy, specific
angular momentum and the radius of the marginally stable orbit, both for null
and timelike geodesics. Then, in the following sections, we particularize these
expressions for solutions written in cylindrical coordinates, oblate spheroidal
coordinates and prolate spheroidal coordinates.

Thus, in section \ref{sec:cyl}, for solutions expressed in cylindrical
coordinates we concluded that all null circular orbits are unstable, as is
illustrated by considering the Chazy-Curzon field. Then, in section
\ref{sec:obl}, we present the oblate spheroidal coordinates and some members of
the family of Morgan-Morgan disks are analyzed. Later, in section \ref{sec:pro},
we consider prolate spheroidal coordinates and the range of stability of the Erez-Rosen
solution is obtained. Finally, results are discussed in section \ref{sec:dis}.

\section{Test particle motion}\label{sec:mot}

The metric for a static axially symmetric spacetime can be written as the Weyl
line element \cite{KSHM},
\begin{equation}
ds^{2} = -e^{2\,\psi}dt^{2} + e^{-2\,\psi}[\rho^{2}d\varphi^{2} +
e^{2\gamma}(d\rho^{2} + dz^{2})], \label{mw}
\end{equation}
where $\gamma$ and $\psi$ are functions of $\rho$ and $z$ only. The ranges of
the coordinates $(\rho, \varphi, z)$ are the usual for cylindrical coordinates
and $-\infty < t < \infty$. The Einstein vacuum equations reduce to the system
of Weyl equations \cite{W1,W2}
\begin{eqnarray}
&&\psi_{,\rho\rho} + \frac{1}{\rho}\psi_{,\rho} + \psi_{,zz} =
0,\label{eq:lap} \\
&&\nonumber    \\
&&\gamma_{,\rho} = \rho\left(\psi_{,\rho}^{2} - \psi_{,z}^{2}\right),
\label{eq:wey1} \\
&&\nonumber    \\
&&\gamma_{,z} = 2\rho\,\psi_{,\rho}\psi_{,z}\label{eq:wey2},
\end{eqnarray}
where (\ref{eq:lap}) is the well-known Laplace equation in cylindrical
coordinates with axial symmetry, which is the integrability condition of the
overdetermined system (\ref{eq:wey1})-(\ref{eq:wey2}).

The corresponding Lagrangian for this line element (\ref{mw}) is given by
\begin{equation}
2{\cal L} = -e^{2 \psi} \dot{t}^{2} + e^{-2 \psi}[\rho^{2} \dot{\varphi}^{2} +
e^{2\gamma}(\dot{\rho}^{2} + \dot{z}^{2})], \label{lagran}
\end{equation}
where the dot represents the derivative with respect to the affine
parameter along the geodesic, $\lambda$. Now, as the Lagrangian is
independent of $t$ and $\varphi$, 
\begin{equation}\label{cmov}
-E = \partial {\cal L}/ \partial \dot{t}, \qquad \ell = \partial {\cal L}/
\partial \dot{\varphi},
\end{equation}
are conserved quantities, where $\ell$ is the specific angular momentum and $E$
is the specific energy with respect to infinity.

From the Lagrangian (\ref{lagran}) we can derive the system of motion equations
\begin{widetext}
\begin{eqnarray}\label{emr}
&&\ddot{\rho} + (\dot{\rho}^{2} - \dot{z}^{2})(\gamma_{,\rho} -
\psi_{,\rho}) + 2\dot{\rho}\dot{z}(\gamma_{,z} - \psi_{,z})  + e^{-2\gamma}\left[E^2\psi_{,\rho}+(\rho \psi_{,\rho} -
1)\frac{\ell^2e^{4\psi}}{\rho^3}\right] = 0, \\
	\nonumber	\\
&&\ddot{z} - (\dot{\rho}^{2} - \dot{z}^{2})(\gamma_{,z} -
\psi_{,z}) + 2\dot{\rho}\dot{z}(\gamma_{,\rho} - \psi_{,\rho}) + e^{-2\gamma}\psi_{,z}\left[E^2+\frac{\ell^2e^{4\psi}}{\rho^2}\right] =
0,\label{emz}
\end{eqnarray}
\end{widetext}
which has a unique solution when conditions $x_{0}^{a} = x^{a}(\lambda_{0})$ and
$u^{a}_{0}= \dot{x}^a (\lambda_{0})$ are given, with $x^a = \rho,z$. The initial
condition for the velocity of the particle is obtained by substitution of
(\ref{cmov}) in (\ref{lagran}).

Now, if we confine the motion of the particle to the equatorial plane $z = 0$, from (\ref{lagran}) we obtain for the radial coordinate $\rho$ the equation
\begin{equation}\label{er}
\dot{\rho}^{2} = e^{-2\gamma} \left[E^{2} - e^{2\psi} \epsilon^{2} -
\frac{\ell^{2}}{\rho^{2}} e^{4\psi} \right],
\end{equation}
with $\epsilon = 1$ for timelike geodesics and $\epsilon = 0$ for null
geodesics. The orbit of the particle in the equatorial plane can be obtained
by solving together the above equation and the equation
\begin{equation}
\dot{\varphi} - \ell e^{2 \psi} \rho^{-2} = 0,
\end{equation}
that follows from (\ref{cmov}). For radial motion we have that $\varphi = \varphi_{0} = constant$, so that $\ell = 0$ in the above equations.

The behavior of the trajectories in the equatorial plane is determined by the
equation (\ref{lagran}), that can be conveniently expressed as
\begin{equation}\label{et}
\frac{\dot{\rho}^{2} e^{2\gamma}}{2} +
\frac{e^{2\,\psi}}{2}\left[\epsilon^{2} + \frac{\ell^{2}}{\rho^{2}}
\,e^{2\,\psi}\right] = \frac{E^{2}}{2},
\end{equation}
so that we can define an {\it effective potential} through 
\begin{equation}\label{pe}
V(\rho) = e^{2 \psi} \left[ \epsilon^{2} + \frac{\ell^{2}}{\rho^{2}} e^{2 \psi}
\right],
\end{equation}
which only depends on $\rho$ and the metric function $\psi$. On the other hand,
in order that the metric (\ref{mw}) be asymptotically flat, the functions $\psi$
and $\gamma$ must vanish at infinity. So, we can obtain the general condition
\begin{equation}\label{copo}
\lim_{\rho\rightarrow\infty} V = \epsilon^{2},
\end{equation}
for all the effective potentials of the form (\ref{pe}).

Now, for circular orbits we have that $\rho = \rho_{c} = constant$ and so
$\dot{\rho} = \ddot{\rho} = 0$. Accordingly, from expression (\ref{et}) follows
that
\begin{equation}\label{ep}
E^{2} = V(\rho),
\end{equation}
with $V(\rho)$ given by (\ref{pe}). Furthermore, the minimums of $V(\rho)$
correspond to stable circular orbits, whereas that the maximums of $V(\rho)$
correspond to unstable circular orbits. So, by computing the derivative of
$V(\rho)$ we obtain the equation for their critical values, which can be written
as
\begin{equation}\label{ma}
\ell^{2} e^{2 \psi}(2 \rho \psi_{,\rho} - 1) + \rho^{3} \epsilon^{2}
\psi_{,\rho} = 0
\end{equation}
and, for the case of null circular orbits ($\epsilon=0$), as
\begin{equation}\label{rn}
2 \rho \psi_{,\rho} - 1 =0.
\end{equation}
So, the radius of the timelike and null circular orbits are given, respectively,
by the roots of the two previous equations. 

The specific angular momentum $\ell$ for massive particles in circular orbits
can be obtained from equation (\ref{ma}) and is given by
\begin{equation}\label{map}
\ell^{2} = \frac{\rho^{3} \psi_{,\rho} e^{- 2 \psi}}{1 - 2 \rho \psi_{,\rho}},
\end{equation}
with the condition $0\leq \rho \psi_{,\rho} \leq 1/2$. So, we can see that the
radius of the circular orbits it depends on $\ell$. Now, by replacing (\ref{map})
into equation (\ref{ep}), we obtain the other constant of motion, $E$, for a
particle moving in a circular trajectory
\begin{equation}\label{epgt}
E^{2} = \frac{e^{2 \psi}(1 - \rho \psi_{,\rho})}{(1 - 2 \rho \psi_{,\rho})},
\end{equation}
where, again, $0 \leq \rho \psi_{,\rho} \leq 1/2$.

The stability condition for circular orbits is given by $V''(\rho_{c}) > 0$. So,
for massless particles the stability condition reduces to
\begin{equation}
\psi_{,\rho} + \rho \psi_{,\rho\rho} > 0,
\end{equation}
whereas that for massive particles with specific angular momentum given by (\ref{map}), the stability condition is given by
\begin{equation}\label{sdp}
\rho \psi_{,\rho\rho} + 3 \psi_{,\rho} + 2 \rho \psi_{,\rho}^{\,2} (2 \rho
\psi_{,\rho} - 3) > 0, 
\end{equation}
with $0 \leq \rho \psi_{,\rho} \leq 1/2$.

Now, one can show that the expressions
\begin{equation}
V''(\rho) = 0
\end{equation}
and
\begin{equation}
\frac{d\ell^{2}}{d\rho} = 0
\end{equation}
are equivalent. Accordingly, the radius of the marginally stable circular orbit can be
obtained through the  two simultaneous equations $V'(\rho) = 0$ and $V''(\rho) = 0$ or
by means of the equation $d\ell^{2}/d\rho = 0$, provided that there exist two critical 
points of the effective potential, one of them corresponding to the stable circular
orbit and the other one to the unstable circular orbit. Thus, the minimum value of
the specific angular momentum as a function of the radius of the circular orbit 
(\ref{map}), it represents the last circular orbit, which is well-known as the
marginally stable circular orbit. For null geodesics the expression is
\begin{equation}\label{romen}
\psi_{,\rho} + \rho\,\psi_{,\rho\rho} = 0,
\end{equation}
and for timelike geodesic is
\begin{equation}\label{romet}
\rho\,\psi_{,\rho\rho} + 3\psi_{,\rho}+
2\rho\,\psi_{,\rho}^{\,2}(2\rho\,\psi_{,\rho} - 3) = 0,
\end{equation}
where $0\leq \rho\,\psi_{,\rho} \leq 1/2$ again.

Finally, we can also find an expression for the angular velocity,
\begin{equation}\label{va1}
\omega = \frac{d\varphi}{dt} = \frac{\dot{\varphi}}{\dot{t}},
\end{equation}
wherein $\dot{t}= E\,e^{-2\,\psi}$ and $\dot{\varphi} =
\ell\,e^{2\,\psi}\rho^{-2}$. For timelike geodesic we obtain
\begin{equation}\label{vagt}
\omega^{2}_{T} = \frac{\ell^{2}e^{8\psi}}{E^{2}\rho^{4}_{c}},
\end{equation}
where $\rho_{c}$ are the roots of equation (\ref{ma}), whereas for null
geodesics we have
\begin{equation}\label{vagn}
\omega_{N} = \frac{e^{2\,\psi}}{\rho_{c}}\,,
\end{equation}
where $\rho_{c}$ are the solutions of expression (\ref{rn}).  The above equations
only depend on the metric function $\psi$, in such a way that the potential $\gamma$
it is not needed for a qualitative analysis of the particle trajectories in Weyl 
spacetimes. However, the function $\gamma$ is necessary for solving the
differential equations of motion of the particle.

\section{Solutions in cylindrical coordinates}\label{sec:cyl}

In {\it spherical coordinates} $(r, \theta)$, the  asymptotically flat solutions of
the equations system (\ref{eq:lap}) - (\ref{eq:wey2}) are \cite{KSHM}
\begin{eqnarray}\label{slcc}
\psi_{k} = &-& \sum_{n = 0}^{k}\frac{C_{n}\,P_{n}}{r^{n+1}}, \\
\nonumber \gamma_{k} = &-& \sum_{l,m = 0}^{k}
\frac{C_{l}C_{m}(l+1)(m+1)}{(l+m+2)r^{l+m+2}} \\
&&\ \times \ (P_{l}P_{m}-P_{l+1}P_{m+1}),
\end{eqnarray}
where $P_{n} = P_{n}(\cos\theta)$ are the usual Legendre polynomials and the $C_{n}$
are constants. Now, in the equatorial plane $z=0$ so that we have
\begin{eqnarray}\label{slcc2}
\psi_{k} = &-& \sum_{n =
0}^{k}C_{2n}\frac{P_{2n}(0)}{\rho^{2n+1}}\,,
\\\nonumber 
\gamma_{k} = &-&
\sum_{l,m = 0}^{k}\frac{C_{2l}C_{2m}(2l+1)(2m+1)}{(2l+2m+2)\rho^{(l+m)/2+1}}\\
&&\ \times \ (P_{2l}P_{2m}-P_{2l+1}P_{2m+1}).
\end{eqnarray}
where $(\rho, z)$ are the usual {\it cylindrical coordinates}, with
$$
\rho=r\sin\theta,\qquad z=r\cos\theta.
$$

We can obtain the radius of a circular null orbit by solving the equation (\ref{rn}),
that in these coordinates reduces to
\begin{equation}\label{rncc}
\sum_{n=0}^{k} \frac{ 2C_{2n}P_{2n}(0)(2n+1) }{\rho^{2n+1}} = 1.
\end{equation}
So, if $C_{2n}P_{2n}(0) > 0$, we obtain
\begin{equation}
\rho ^{2 k+1}-\sum _{n=0}^k  a_{2 n}\rho ^{2 n} = 0,
\end{equation}
a polynomial in $\rho$ off odd order that has, at least, a real root. Moreover,
since there is only one change in sign, there is a positive root. Accordingly, we
can conclude that there exist null circular orbits. On the other hand, it is easy to
see that the stability condition it is not satisfied, since
\begin{equation}
\rho\,\psi_{k,\rho\rho} + \psi_{k,\rho}=
-\sum_{n=0}^{k} \frac{ C_{2n}P_{2n}(0)(2n+1)^{2} }{\rho^{2n+2}} < 0
\end{equation}
and so all the circular orbits are unstable. Finally, we can ask for the existence of
a marginally stable circular orbit, which must satisfy $\rho\,\psi_{,\rho\rho} +
\psi_{,\rho}=0$, but we find that there are not positive roots as the
corresponding polynomial has not any change of sign. Therefore, there
is not any marginally stable orbit.

Now, in order to illustrate the above considerations, we take the simplest case of 
the family (\ref{slcc}), the Chazy-Curzon solution \cite{Chazy,CU},
\begin{equation}\label{chacur}
\psi_0 = - \frac{m}{r},\qquad\gamma_0 = - \frac{m^2 \sin^2 \theta}{2 r^2},
\end{equation}
which can be obtained taking $k=0$ and $C_{0} = m > 0$ in (\ref{slcc}). As we can see,
although the metric function $\psi_0$ is spherically symmetric, the full solution 
(\ref{chacur}) is not. In the equatorial plane, the metric functions reduce to
\begin{equation}
\psi_0 = - \frac{m}{\rho},\qquad\gamma_0 = - \frac{m^2}{2\rho^2}.
\end{equation}
So, according to (\ref{rncc}), for this solution the radius of the unstable circular 
orbit is $\rho = 2m$. Then, with $\rho = 2m$, the values of the specific energy and the
angular velocity of the particle are
\begin{equation}
\omega_{N} = (2me)^{-1},\qquad E_{N} = \ell\,\omega_{N},
\end{equation}
where $\ell$ is an arbitrary constant.

\begin{figure*}
\includegraphics[width=4.5in]{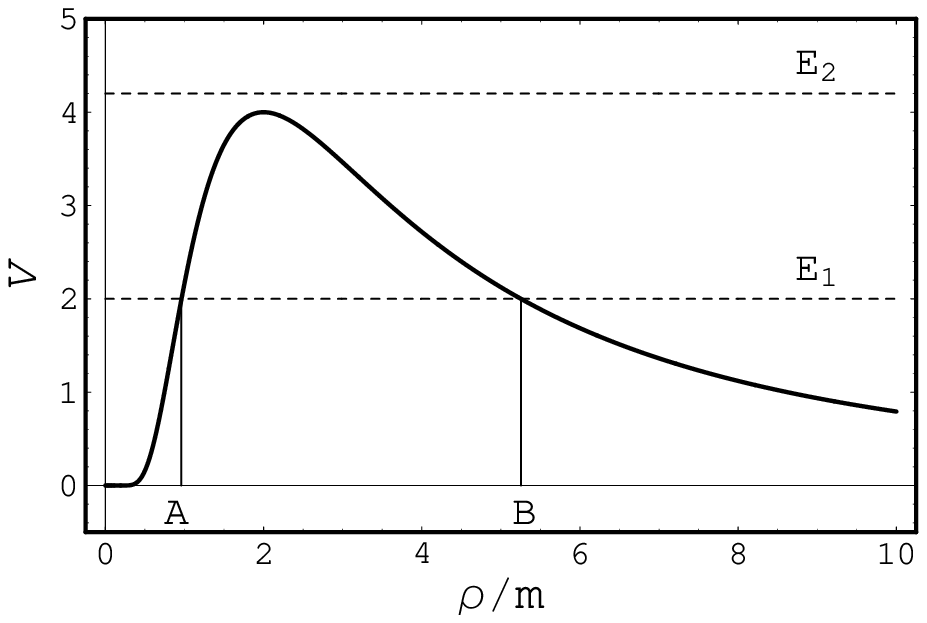}
\caption{The effective potential for the massless particle in the Chazy-Curzon
field. Here we take $\ell=4me$.}\label{pccn}
\end{figure*}

In Fig. \ref{pccn} we show the effective potential for lightlike geodesics in
the Chazy-Curzon solution. As we can see, there is a maximum at $\rho = 2m$ with a
value of $(2me)^{-2}\ell^{2}$. Trajectories can be described using the horizontal
lines ($V = E^{2}$) to different values of the quantities $m$ and $\ell$.  When $0 < 
E_{1} < (2me)^{-2}\ell^{2}$ the motion corresponds to a particle with specific energy 
$E_{1}$ coming from infinity until reach the turning point $B$ and then going 
back to infinity. There is also a potential well, when the motion it is confined to $0< 
\rho/m < A$, and a not allowed region, for $A < \rho/m <B$. On the other hand, when the 
specific energy it is greatest than $(2me)^{-2}\ell^{2}$ as in the horizontal line 
$E_{2}$, there are no turning points and the particle it moves only in one direction. 
Now, although we do not consider here solutions with other values of $k$ in 
(\ref{slcc2}), it can be shown that in general all the effective potentials behave as 
depicted in Fig. \ref{pccn}, whenever the condition $C_{2n}P_{2n}(0) > 0$ be assumed in 
this family.

On the other hand, for timelike geodesics the specific energy and specific angular 
momentum of a particle in a circular orbit are, respectively,
\begin{equation}\label{mocc}
\ell^{2}=\frac{\rho^{2}m\,e^{2m/\rho}}{\rho-2m},\qquad E^{2}=e^{-2m/\rho},
\end{equation}
where $\rho\geq2m$, so that
\begin{equation}
\omega_{T}^{2} = \frac{e^{-4 m/\rho } m}{(\rho -2 m) \rho ^2}.
\end{equation}
The radius of the marginally stable circular orbit can be obtained by solving the 
equation (\ref{romet}) and so we can find the  corresponding specific angular momentum
by replacing this radius in equation (\ref{map}). For the Chazy-Curzon field we have
\begin{eqnarray}
&&\rho = 5.23m,\qquad \ell=3.52m,\\
&&E=0.826,\qquad \omega_{T}^{2} =\frac{0.00525}{m^2},
\end{eqnarray}
for the marginally stable circular orbit.

\begin{figure*}
\includegraphics[width=4.5in]{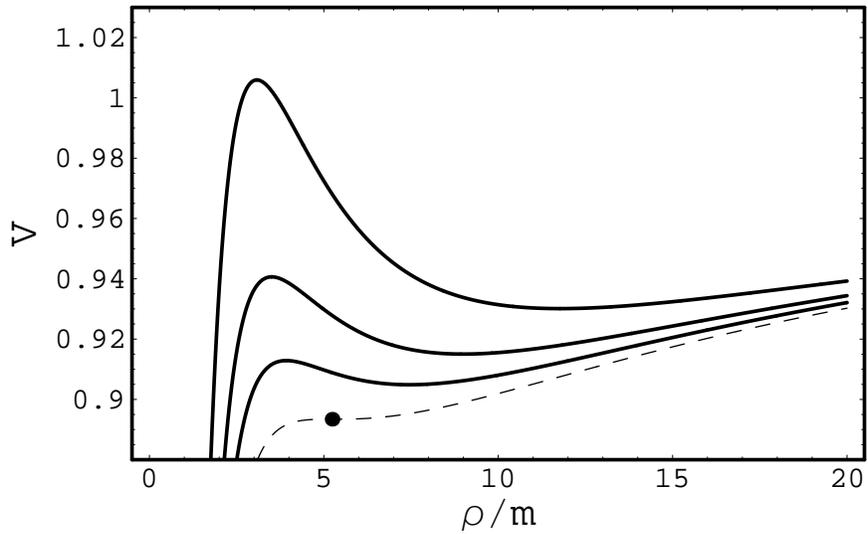}\\
\caption{Effective potential as a function of $\rho/m$ for timelike geodesics, with 
different values of $\ell/m$, for the Chazy-Curzon field. For the curves from top to 
bottom we take $\ell/m=4.1,3.8,3.65$ and $\ell=3.52m$, respectively. The point in the 
dotted curve corresponds to $\rho/m=5.23$, which is the radius of the marginally
stable orbit with $\ell/m=3.52$.}\label{pcct1}
\end{figure*}

\begin{figure*}
\includegraphics[width=4.5in]{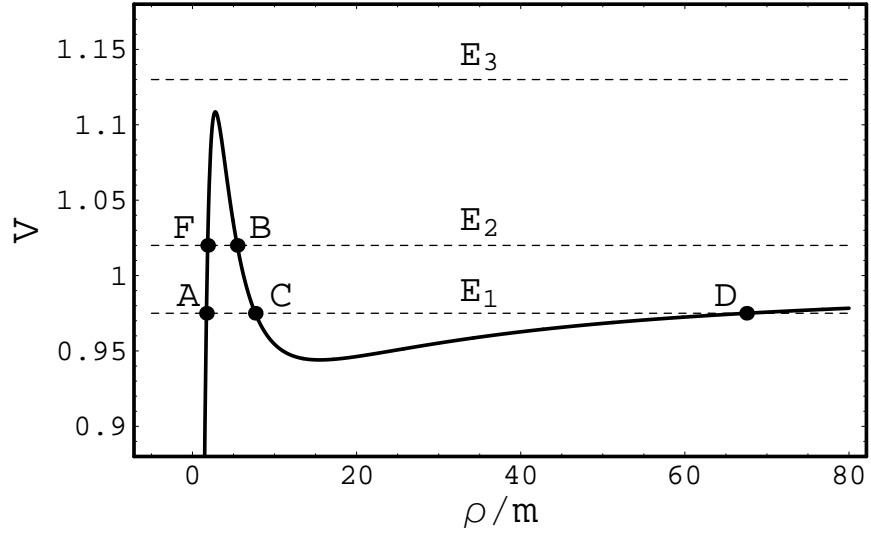}\\
\caption{Effective potential for timelike geodesic with $\ell=4.5m$ in
the Chazy- Curzon spacetimes.}\label{pcct3}
\end{figure*}

\begin{center}
\begin{figure*}
\includegraphics[width=4.5in]{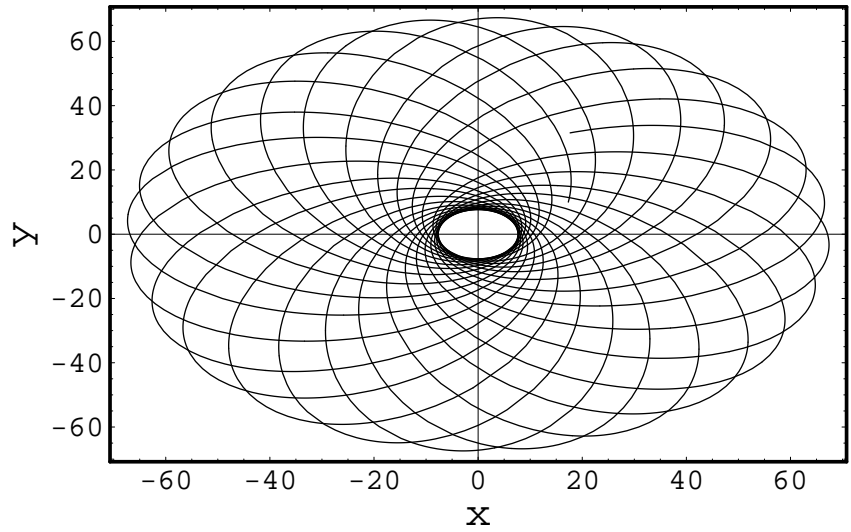}
\caption{Orbit of the particle corresponding to the effective
potential of the Fig. \ref{pcct3} with $E_{1} = 0.975$. This orbit
is bounded between $C \approx 7.66$ and $D \approx67.52$. The initial
conditions are $\dot{\rho}(t=0) \approx 0.17$, $\varphi(t=0) = \pi/6$ and $
\rho(t=0)=20$.}\label{occ1}
\end{figure*}
\end{center}

\begin{center}
\begin{figure*}
\includegraphics[width=4.5in]{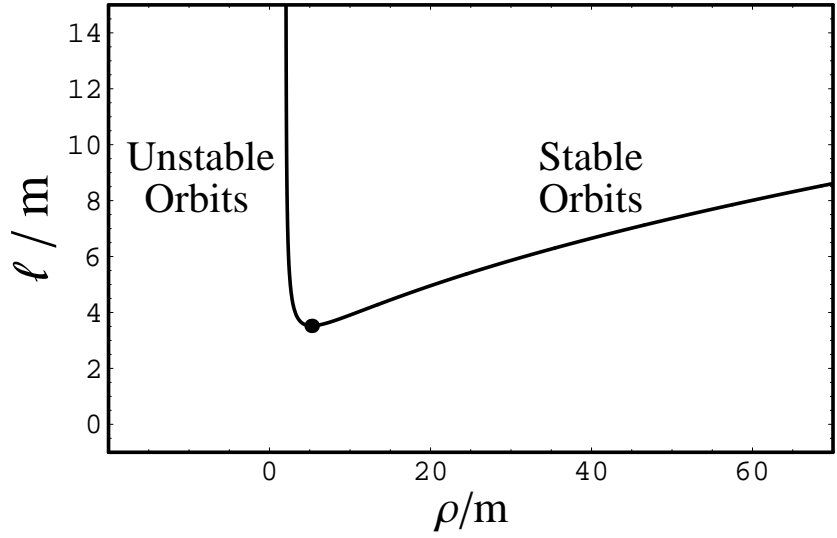}
\caption{The specific angular momentum, $\ell/m$, as a function of the radius of the 
circular orbit, $\rho/m$, for timelike geodesic in the Chazy-Curzon field. The point have 
coordinates $(\ell/m,\rho/m) = (3.52,5.23)$. The range of stability is $3.52m \leq \ell < 
\infty$ and $5.23m \leq \rho < \infty$.}\label{fig:mocc}
\end{figure*}
\end{center}

The graphics of the effective potential (\ref{pe}) for timelike geodesics in the
Chazy-Curzon spacetime are presented in Figs. \ref{pcct1} and \ref{pcct3}. In
Fig. \ref{pcct1} we can see that the shape of the  effective potential curve depends
only on the angular momentum $\ell$ and that all the curves have two circular orbits, 
except the doted curve. In this graph the doted curve has $\ell = 3.52m$ and represents 
the marginally stable circular orbit. In the Fig. \ref{pcct3} we show the potential with
$\ell=4.5m$ and, with the dotted lines, different values of the energy $E^2$ in order 
to analyze the possible orbits. Thus, for $E_{1}$ we have three different radius as the 
horizontal line cut the potential in three points. For the least radius $A$, the particle 
is confined in potential well as in Fig. \ref{pccn}, $0<\rho/m<A$. Whereas for the others 
radius, C and D, we obtain a bounded orbit between these radius, i.e., if we take $E_{1} 
= 0.975$, we obtain a bounded orbit between $C = \rho/m \approx 7.66$ (perihelio) and $D 
=  \rho/m\approx 67.52$ (aphelio),  Fig. \ref{occ1}. When the energy is  
$E_{2}$ we get a turning point, so for $E_{2}=1.02$ the turning point is $B=\rho/m
\approx5.45$, we show in Fig. \ref{occ2} the corresponding graph. For the point $F$ 
with energy $E_{2}=1.02$, the particle is confined in the potential well, as in $A$. 
Whereas for an energy $E_{3}$, greater then the maximum of the potential, there 
are not turning points and the particle moves only in a direction. Finally, in the Fig. 
\ref{fig:mocc} we present the range of stability for particles moving in a circular orbit 
by plotting the specific angular momentum (\ref{mocc}) as a function of the radius of
the circular orbit. The range of stability is $3.52m \leq \ell < \infty$ and $5.23m
\leq \rho < \infty$. The point have coordinates $(\ell/m,\rho/m) = (3.52,5.23)$, and 
corresponds to the radius of the marginally stable circular orbit.

\section{Solutions in Oblate Spheroidal Coordinates}\label{sec:obl}

In the {\it oblate spheroidal coordinates}, the solution of the Laplace equation
(\ref{eq:lap}) is
\begin{equation}\label{mam}
\psi_{n}= -\sum_{k = 0}^{n}C_{2k}P_{2k}(\eta)i^{2k + 1}Q_{2k}(i\xi),
\end{equation}
where $C_{2k}$ are constants, $P_{k}$ are the Legendre polynomials
and $Q_{k}$ are the Legendre functions of second kind \cite{BA}. This solution
repesents the exterior Newtonian potential for an infinite family of axially symmetric 
finite thin disks, recently studied by Gonz\'alez and Reina \cite{GR1}, and whose first 
member, $n=0$, is the well-known Kalnajs disk \cite{KAL}. We also studied in another 
paper \cite{RGS} the kinematics around the first four members of this family by means of 
the Poincar\'e surfaces of section and Lyapunov characteristic numbers and
fond chaos in the case of disk-crossing orbits and completely regular motion in
other cases.

The constants $C_{2k}$ appearing in (\ref{mam}) are given by
$$
C_{2k}= \frac{m G}{2 a} \left[\frac{\pi^{1/2} (4k+1) (2n+1)!}{2^{2n} (2k+1) (n - k)! \Gamma(n +
k + \frac{3}{2} ) q_{2k+1}(0)} \right],
$$
where $q_{2k} (\xi) = i^{2k+1} Q_{2k} (i\xi)$, $m$ is the mass of the disk  and $G$ is the gravitational
constant. Now, due to the presence of the term $(n - k)!$ at the denominator,
all the $C_{2k}$ constants vanish for $k > n$. The variables $\eta$ and $\xi$ are
the oblate coordinates related with the cylindrical coordinates by
\begin{equation}\label{cco}
\rho^{2} = a^{2}(1 + \xi^{2})(1 - \eta^{2}),\qquad z =a\xi\eta,
\end{equation}
where $a$ is a constant, $-1 \leq \eta\leq 1$ and $0 \leq \xi < \infty$.
In (\ref{cco}) $a$ is the radius of the thin disk, henceforth we
take $a = 1$. In the plane $z = 0$, we have two regions: if $\xi =0$ then $\eta = 
\sqrt{1-\rho^{2}}$, whereas if $\eta=0$ then $\xi = \sqrt{\rho^{2}-1}$. These two
regions correspond to the regions inside and outside of the disk, respectively.

We now write the different equations for this two regions in the simple case of
null geodesics. So, from (\ref{rn}), the radius of the circular orbits inside the
source is
\begin{equation}\label{rnco}
\sum_{k = 1}^{n}4kC_{2k}q_{2k}(0)\left[P_{2k-1}(\eta) -
\eta P_{2k}(\eta)\right] = \eta,
\end{equation}
where $\eta = \sqrt{1-\rho^{2}}$. The stability condition in oblate
spheroidal coordinates, inside of the disk, take the form
\begin{equation}
\eta(1-\eta^{2})\psi_{n,\,\eta\eta} - (1+\eta^{2})\psi_{n,\,\eta}\leq 0,
\end{equation}
that, for the solution (\ref{mam}), can be written as
\begin{equation}
\sum_{k = 1}^{m}2kq_{2k}(0)\left[2k\eta P_{2k}(\eta) +
P_{2k-1}(\eta)\right]>0.
\end{equation}
Now, for null geodesics outside of the disk, the radius of the circular orbit
can be obtained from the expression
\begin{equation}
\sum_{k = 0}^{n}2C_{2k}P_{2k}(0)\left[\xi q_{2k}(\xi) + q_{2k+1}(\xi)
\right] + \xi = 0,
\end{equation}
where $\xi=  \sqrt{\rho^{2}-1}$. The corresponding
stability condition is
\begin{equation}
\xi(1+\xi^{2})\psi_{,\xi\xi} + (\xi^{2}-1)\psi_{,\xi}\leq0,
\end{equation}
that using (\ref{mam}), becomes
\begin{equation}
-\sum_{k = 0}^{n}C_{2k}(2k+1)\left[2\xi q_{2k}(\xi) +
q_{2k+1}(\xi)\right]>0.
\end{equation}

The first solution of (\ref{mam}), when $n=0$, was obtained
independently by Zipoy \cite{ZI} and Vorhees \cite{VO}, and
interpreted by Bonnor and Sackfield \cite{BO} as the gravitational
field of a pressureless static thin disk, this disk is singular
at the rim. The function $\psi$ for the first three members of family of disks 
(\ref{mam}) is given by \cite{GR1,RGS}
\begin{eqnarray}
\psi_{1} &=& - \mu [ \cot^{-1}\xi  + A (3\eta^{2} - 1)],
\label{eq:4.22}   \\
\psi_{2} &=& - \mu [ \cot^{-1} \xi + \frac{10 A}{7}
(3\eta^{2} - 1) \nonumber \\
 &&+ \ B ( 35 \eta^{4} - 30 \eta^{2} + 3)], \label{eq:4.23}  \\
\psi_{3} &=& - \mu[ \cot^{-1} \xi + \frac{10 A}{6} (3
\eta^{2} - 1)  \nonumber \\
    && + \ \frac{21 B}{11} (35 \eta^{4} - 30 \eta^{2} + 3)
\nonumber   \\
&& + \ C (231 \eta^{6} - 315 \eta^{4} + 105 \eta^{2} - 5) ],
\label{eq:4.24}
\end{eqnarray} 
with
 \begin{eqnarray*}
A &=& \frac{1}{4} [(3\xi^{2} + 1) \cot^{-1} \xi - 3 \xi ],
\\
B &=& \frac{3}{448} [ (35 \xi^{4} + 30 \xi^{2} + 3) \cot^{-1} \xi
- 35 \xi^{3} - \frac{55}{3} \xi ], \\
C &=& \frac{5}{8448} [ (231 \xi^{6} + 315 \xi^{4} + 105 \xi^{2} +
5) \cot^{-1} \xi  \nonumber \\
\quad \quad &&- 231 \xi^{5} - 238 \xi^{3} - \frac{231}{5} \xi ],
\end{eqnarray*} 
wher $\mu=m/a$ and $n=1,2,3$, respectively.

For $n=1$, the other metric function is \cite{KSHM,M1,M2}
\begin{eqnarray}
\gamma_{1} &=& 9 \mu^{2} (\eta^{2} - 1) \left[ 9 \xi^{2} \eta^{2} -\xi^{2}
+ 4 \eta^{2} + 4 \right.\nonumber\\
&& + (\xi^{2} + 1) (9 \xi^{2} \eta^{2} - \xi^{2} +
\eta^{2}-1) (\cot^{-1}\xi)^{2} \nonumber\\
&&\left. - 2 \xi (9 \xi^{2} \eta^{2} - \xi^{2} + 7 \eta^{2} + 1) \cot^{-1}\xi \right]/16.
\end{eqnarray}
This disk is also singular at the rim \cite{S1}. For $n = 2$, the metric function
$\gamma$ is given by
\begin{eqnarray}\label{ns} \nonumber
\gamma_{2} &=& 25\mu^{2}(\eta ^2-1) \left\{9\xi ^6 \left(1225 \eta ^6-1275 \eta^4+315
   \eta^2 - 9\right) \right.\\ \nonumber
&&\left. + 3 \eta ^4 \left(5050 \eta ^6-3630 \eta ^4+366 \eta^2 + 6\right)\xi^4 \right.\\ \nonumber
&& + \left(4945 \eta^6 - 723 - 45 \eta ^2-81\right) \xi^2 \\ \nonumber
&& - 6 \left[375 \eta^6 + 113 \eta^4 + 15 \eta^2 + 27\right. \\ \nonumber
&& + 3 \left(1225 \eta^6 - 1275 \eta^4 + 315 \eta^2 - 9\right) \xi^6 \\ \nonumber
&& + \left(6275 \eta^6 - 4905 \eta^4 + 681 \eta^2 - 3 \right) \xi^4 \\ \nonumber
&& \left. + \left(3005 \eta^6 - 1111 \eta^4 - 105 \eta^2 + 3 \right) \xi^2
\right]\xi \cot^{-1}\xi \\ \nonumber
&& + 9 (\xi^2 + 1) \left[9 \eta^6 + 5 \eta^4 - 5 \eta^2 - 9 \right. \\ \nonumber
&& + \xi^6 \left(1225 \eta^6 - 1275 \eta^4 + 315 \eta^2 - 9 \right) \\ \nonumber
&& + \xi^4 \left(1275 \eta^6 - 785 \eta^4 + 17 \eta^2 + 5 \right) \\ \nonumber
&&\left. + \xi^2 (315 \eta^6 - 17 \eta^4 - 47 \eta^2 + 5) \right]
   (\cot ^{-1}\xi)^2 \\
&&\left. + 256 \left(\eta^6 + \eta^4 + \eta^2 + 1 \right) \right\}/ 2048.
\end{eqnarray}
For $n\geq2$, although the metric function $\gamma$ can be easily obtained by
integrating the equations (\ref{eq:wey1}) - (\ref{eq:wey2}) properly written in oblate 
spheroidal coordinates, they are not explicitely presented here due to their higly 
involved expressions.

Now we analyze some examples. If we take $n=1$, the second member of family of disks,
we obtain for the radius of circular orbit, the angular velocity and specific
energy corresponding to this radius, the relations
\begin{equation}
\rho^{2} = \frac{2}{3\pi\mu},\qquad \omega_{N}^{2} = \frac{E^{2}}{\ell^{2}} = \frac{3\pi  \mu}{2}\, e^{1-3 \pi  \mu },
\end{equation}
where $\mu \geq 2/3\pi$, $\ell$ is an arbitrary constant and the radius corresponds to
a stable equilibrium of the effective potential for a null geodesic. Now, outside of the 
source the effective potential increases until a maximum value, corresponding to the 
unstable circular orbit, and then diminishes until $0$ when $\rho$ increases, according 
to (\ref{copo}). Again, we cannot find the marginally stable circular orbit in this disk 
for the massless particles.

\begin{center}
\begin{figure*}
\includegraphics[width=4.5in]{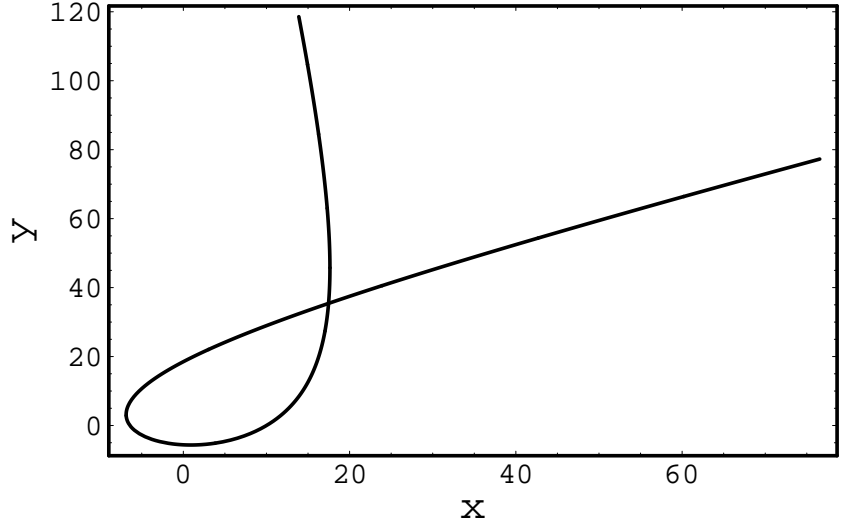}
\caption{Orbit of the particle corresponding to the effective
potential of the Fig. \ref{pcct3} with $E_{2} = 1.02$. The initial
conditions are $\dot{\rho}(t=0) \approx 0.17$, $\varphi = \pi/6$ and $\rho=20$.
}\label{occ2}
\end{figure*}
\end{center}

\begin{center}
\begin{figure*}
\includegraphics[width=4.5in]{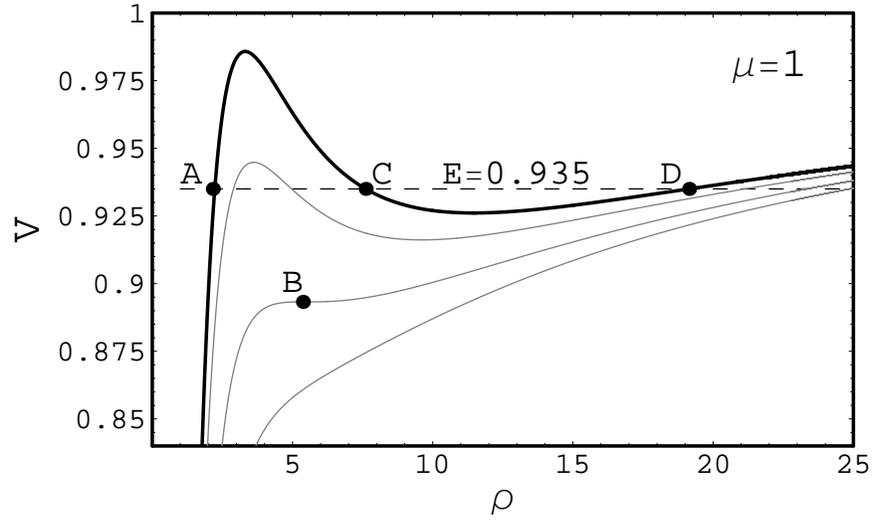}
\caption{Effective potential outside of source for the third member of the family of 
Morgan and Morgan disks, $n=2$, for a timelike geodesic with $\mu=1$.}\label{ptmm1}
\end{figure*}
\end{center}

\begin{center}
\begin{figure*}
\includegraphics[width=4.5in]{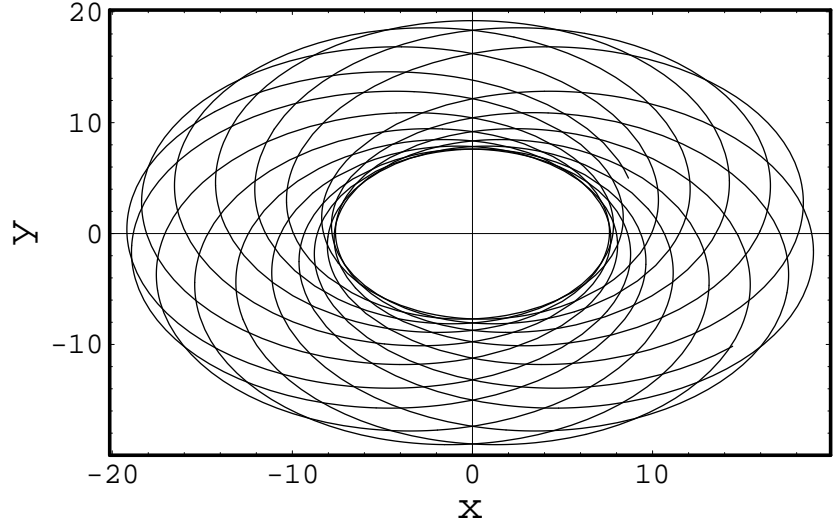}
\caption{Orbit of a particle with $E = 0.935$. The initial conditions are $\dot{\rho}(0) 
= 0.0903$, $\varphi(0) = \pi/6$ and $\rho(0)=10$. The trajectory is between $C\leq\rho
\leq D$.}\label{otmm2}
\end{figure*}
\end{center}

The behavior of the effective potential for different values of parameters is similar to 
that we displayed in Fig. \ref{occ2}, corresponding to the timelike geodesic. In the 
effective potential of the Fig. \ref{occ2} we put $\ell =8$, 6 (gray curve) and 3.56 
(dotted curve). The behavior for different values of the parameters is similar, so we 
take $\mu=1$ as an example. In the exterior case we consider the third member of the 
family, $n=2$, Fig. \ref{ptmm1}. In this graph the point $B$ corresponds to the radius of 
the marginally stable circular orbit, which have $\ell \approx 3.688$. The points $C$ and 
$D$ have angular momentum $\ell = 4.2$ and effective potential $0.935$, the motion is 
bounded between the radius $7.61$ and $19.14$, Fig. \ref{otmm2}. In this graph we choose 
$\mu =1$.

\section{Solutions in Prolate Spheroidal Coordinates}\label{sec:pro}

The general static axisymmetric vacuum solution for $\psi$ in {\it prolate
spheroidal coordinates} $(u,v)$ is given by \cite{Q1}
\begin{equation}\label{slcp}
\psi_{l} = \sum_{n=0}^{l}(-1)^{n+1}d_{n}Q_{n}(u)P_{n}(v),
\end{equation}
where $u \geq 1$, $-1\leq v \leq1$, and the $d_{n}$ are constants related
with the multipole moments \cite{Q1,MA}. $P_{n}$ are
the Legendre polynomials and $Q_{n}$ are Legendre functions of
second kind. These coordinates are related with Weyl's canonical
coordinates by
\begin{equation}
\rho^{2} = m^{2}(1-v^{2})(u^2-1),\qquad z=m uv,
\end{equation}
being $m$ the mass of the source that produce the field.

The asymptotically flat solution for $\gamma$ was found for Quevedo in \cite{Q1}. The 
monopolar solution, $l = 0$, with $d_{0} = 1$ corresponds to the Schwarzschild field.
The solution of equation (\ref{rn}) for the monopolar solution is $u = 2$, that is the 
unstable radius of a null circular orbit in the Schwarzschild field. For a complete
study of motion in Schwarzschild field in the equatorial plane, see \cite{V1}. The 
solution of (\ref{slcp}) for $l = 2$ is the Erez-Rosen metric \cite{ER},
\begin{eqnarray}\label{per}
\psi_{1} &=& \frac{d_{0}}{2}\ln \frac{u-1}{u+1}\\
&&+\nonumber\frac{d_{2}}{2}(3v^{2}-1)\left[\frac{1}{4}(3u^{2}-1)\ln
\frac{u-1}{u+1} + \frac{3}{2}u\right].
\end{eqnarray}
In this solution, $d_{0}$ and $d_{2}$ are related with the monopole and
arbitrary quadrupole moment, respectively \cite{MA}. The study of orbits in this
solution was developed by different authors 

Now, in this section we expose the expressions for the different quantities 
corresponding to the motion of a particle in a circular orbit in prolate spheroidal 
coordinates. That is, the specific energy, the angular velocity and the radius of the 
marginally stable circular orbit, which are obtained through of the effective potential 
in the equatorial plane,
\[
V = e^{2 \psi_{l}} \left(\epsilon
   ^2+\frac{e^{2 \psi_{l}}
   \ell^{2}}{\left(u^2-1\right)
   m^2}\right).
\]
We begin with the equations corresponding to null geodesics, when $\epsilon = 0$. So,
the radius of the circular orbits can be determinate by
\begin{equation}
2(u^{2}-1)\psi_{l,\,u} = u,
\end{equation}
that in terms of (\ref{slcp}) takes the form
\begin{equation}\label{rncp}
\sum_{n=0}^{l}2d_{2n}P_{2n}(0)(u^{2} - 1)Q'_{2n}(u)=u,
\end{equation}
with the stability condition
\begin{equation}
u(u^{2}-1)\psi_{,uu} + (u^{2}+1)\psi_{,u}\geq0,
\end{equation}
that using (\ref{slcp}) can be written as
\begin{widetext}\begin{equation}
\sum_{n=0}^{l} d_{2n} P_{2n}(0) \left[ (u^{2} - 1) Q'_{2n}(u) - 2 n u (2n + 1)
Q_{2n}(u) \right] \geq 0,
\end{equation}\end{widetext}
where the equality corresponds to the equation for the radius of the marginally stable 
circular orbit. Finally, the other expressions are found by means of (\ref{ep}) and 
(\ref{vagn}).

On the other hand, for the massive particle we obatin the expressions
\begin{eqnarray}\label{mocp}
\ell^{2} &=& \frac{e^{-2 \psi}
   \left(u^2-1\right)^2
    m ^2 \psi
   _{,u}}{u-2 \left(u^2-1\right)
   \psi_{,u}},\\
   &&	\nonumber	\\
E^{2} &=& \frac{e^{2 \psi}  \left[u-(u^2-1)
   \psi_{,u}\right]}{u-2 \left(u^2-1\right)
   \psi_{,u}},
\end{eqnarray}
where $u-2 \left(u^2-1\right)\psi_{,u}\geq0$ in order that the energy per mass unit and
the angular moment per mass both be not imaginary. The stability condition and the
radius of the marginally stable circular orbit are
\begin{eqnarray}\nonumber
&&\vspace{-1cm}\psi_{,u}\left\{3u^2 + 2 + 2(u^2-1) \psi_{,u}
\left[2 (u^2 - 1) \psi_{,u} - 3 u \right] \right\} \\
&&\ + \ u (u^2 - 1) \psi_{,uu} \geq 0,
\end{eqnarray}
with $u-2 \left(u^2-1\right)\psi_{,u}\geq0$.

\begin{center}
\begin{figure*}[h]
\includegraphics[width=4.5in]{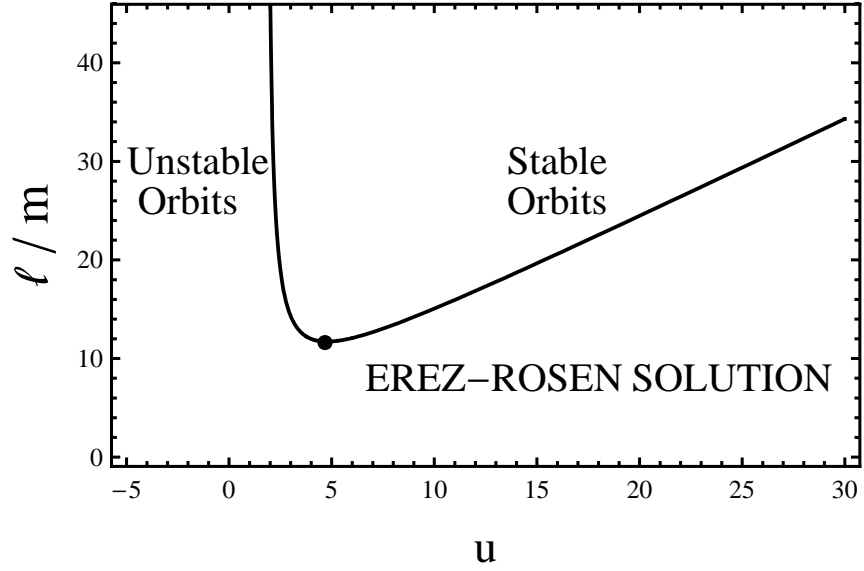}
\caption{Specific angular momentum, $\ell/m$, as a function of $u$ for the circular
orbit in the Erez-Rosen field with $d_{0} = 1$ and $d_{2} = 1.5$. }\label{moer}
\end{figure*}
\end{center}

Finally, we show in Fig. \ref{moer}the region of stability by means of a graph of the specific angular moment (\ref{mocp}). In particular, by taking arbitrary
values of the parameters in the Erez-Rosen solution, one can see that the range is
\[
0\leq u \leq 4.77, \qquad 4.77\leq u < \infty,
\]
and $11.62 \leq \ell < \infty$ for stable orbits, where we choose $d_{0} = 1$ and
$d_{2} = 1.5$. Here the marginally stable circular orbit have coordinates $(\ell/m,u)
= (11.62,4.77)$.

\section{Concluding remarks}\label{sec:dis}

In this paper we analyzed the behavior of free test particles in the equatorial plane
of static axisymmetric spacetimes. We presented several general expressions for the 
circular orbit in null and timelike geodesics: radius, specific energy, specific angular 
momentum, angular velocity and radius of the marginally stable circular orbit, all of 
them obtained through of an effective potential. The specific angular momentum was 
presented for the timelike geodesic and was used to determine the range of stability of 
the orbit of the particle, so the the minimum value represents the marginally stable 
circular motion.

In order to find the trajectory of the particle, we analyzed the analytical results 
obtained. The character of the motion is determined essentially by means of
the behavior of the effective potential. Thus, we displayed different graphs of effective 
potential before resolve the differential equations of motion of the particle. Then, we 
began with the Chazy-Curson field in the case of the cylindrical coordinates, as 
discussed in section III. The motion of particles around oblate deformed bodies was 
developed in section IV, by means of the analysis of the properties of some
member of the family disks (\ref{mam}). On the other hand, the prolate case was presented 
in section V, where we found the range of stability of the Erez-Rosen solution in the 
special case of mass particles.

In summary, we concluded that for these solutions all the circular orbits are unstable 
only in the case of null geodesic, whereas that do not exist marginally stable circular 
orbit for null geodesic. In contrast, we found that for mass particles the orbits
can be unbounded, bounded or circular. This behavior can be seen by means of the
effective potential and verified through the numerical solution of the equations
of motion. Moreover, for the timelike geodesic we found the radius of the marginally 
stable circular orbit in different coordinate systems, cylindrical, prolate and oblate.

\begin{acknowledgments}
F. L-S want to thank the financial support from {\it Vicerrector\'ia Acad\'emica}, 
Universidad Industrial de Santander.
\end{acknowledgments}

\newpage

\begin{center}
\begin{figure*}
\includegraphics[width=4.5in]{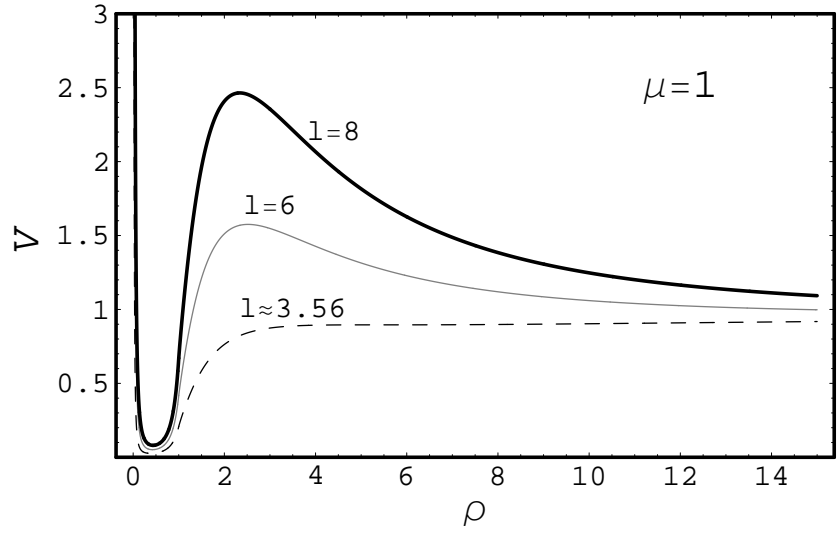}
\caption{Effective potential inside and outside of  the source for mass particles for
the second member of the family of Morgan and Morgan disks, $n=1$, with $\mu=1$.}
\end{figure*}
\end{center}


\end{document}